\documentclass{aastex}
\usepackage{emulateapj5,times,mathptm}
\slugcomment{Accepted for the publication in the {\it Astrophysical Journal}}
\def\farcm{\hbox{$.\mkern-4mu^\prime$}}
\def\farcs{\hbox{$.\mkern-4mu^{\prime\prime}$}}

\def\la{\mathrel{\hbox{\rlap{\hbox{\lower4pt\hbox{$\sim$}}}\hbox{$<$}}}}
\def\ga{\mathrel{\hbox{\rlap{\hbox{\lower4pt\hbox{$\sim$}}}\hbox{$>$}}}}

\shortauthors{Park}
\shorttitle{N49}

\begin{document}
\title{X-Ray Emission from Multi-Phase Shock in the Large Magellanic 
Cloud Supernova Remnant N49}

\author{Sangwook Park\altaffilmark{1}, David N. Burrows, Gordon P. 
Garmire, and John A. Nousek}

\affil{Department of Astronomy and Astrophysics, Pennsylvania State
University, 525 Davey Laboratory, University Park, PA. 16802}


\author{John P. Hughes}

\affil{Department of Physics and Astronomy, Rutgers University,
136 Frelinghuysen Road, Piscataway, NJ. 08854-8109}


\author{and}

\author{Rosa Murphy Williams}

\affil{Department of Astronomy, B619E - LGRT, University of Massachusetts,
Amherst, MA 01003}






\altaffiltext{1}{park@astro.psu.edu}

\begin{abstract}

The supernova remnant (SNR) N49 in the Large Magellanic Cloud 
has been observed with the Advanced CCD Imaging Spectrometer 
(ACIS) on board the {\it Chandra X-Ray Observatory}. The superb 
angular resolution of the {\it Chandra}/ACIS images resolves a
point source, the likely X-ray counterpart of soft
gamma-ray repeater SGR 0526$-$66, and the diffuse filaments 
and knots across the SNR. These filamentary features represent 
the blast wave sweeping through the ambient interstellar medium 
and nearby dense molecular clouds. We detect metal-rich ejecta 
beyond the main blast wave shock boundary in the southwest of the 
SNR, which appear to be explosion fragments or ``bullets'' 
ejected from the progenitor star. The detection of strong H-like 
Si line emission in the eastern side of the SNR requires multi-phase 
shocks in order to describe the observed X-ray spectrum, whereas 
such a multi-phase plasma is not evident in the western side. This 
complex spectral structure of N49 suggests that the postshock 
regions toward the east of the SNR might have been reheated by the 
reverse shock off the dense molecular clouds while the blast 
wave shock front has decelerated as it propagates into the dense 
clouds. The X-ray spectrum of the detected point-like source is 
continuum dominated and can be described with a power law of 
$\Gamma$ $\sim$ 3. This provides a confirmation that this point-like 
X-ray source is the counterpart of SGR 0526$-$66 in the quiescent state.

\end{abstract}


\section {\label {sec:intro} INTRODUCTION}

N49 is a bright supernova remnant (SNR) in the Large Magellanic 
Cloud (LMC) \citep{dopita79,long81}. The low metal abundances 
of N49 \citep{danziger85,russell90} indicated a substantial 
mixture of the supernova (SN) ejecta material with the LMC 
interstellar medium (ISM), which is consistent with the relatively 
old dynamical age as derived to be $\sim$5000 $-$ 10000 yr 
\citep{long81,vancura92}. The most remarkable feature of N49 is 
strong enhancements of the optical/X-ray surface brightness in 
the south-east.  This asymmetric intensity distribution has been 
attributed to the interaction of the blast wave with the nearby 
molecular clouds located in the east of the SNR 
\citep{vancura92,banas97}. The radio images show similar 
intensity distribution to the X-ray morphology, but with 
a complete shell in the west as well as in the east 
\citep{dickel95}. The radio spectral index of $\alpha$ = $-$0.58 
\citep{dickel98} is typical for a shell-type SNR. Although the 
mean fractional polarization over the whole remnant is low ($<$5\%), 
the western side of the SNR is highly polarized. The highly 
organized magnetic field in the west of the SNR is supportive of 
the less-disturbed blast wave expanding into the low density ISM 
in the west \citep{dickel98}. 

Vancura et al. (1992) have performed extensive optical/UV spectroscopy 
and have found that N49 is interacting with dense clouds ($n_H$ $\sim$ 150 
cm$^{-3}$ on average) while the preshock intercloud density is $\sim$0.9 
cm$^{-3}$. This dense ISM is most likely associated with the CO clouds 
detected in the east of the SNR \citep{banas97}. The bright optical/UV 
emission arises from ``sheets'' of emission generated by slow 
radiative shocks of $v$ $\leq$ 140 km s$^{-1}$ as the blast wave interacts 
with the dense ambient material, while the X-ray emission appears from 
the fast shock ($v$ $\sim$ 730 km s$^{-1}$) propagating in the intercloud 
medium. 

The X-ray spectrum of N49 as obtained with {\it ASCA} was fitted with a 
non-equilibrium ionization (NEI) model with $kT$ $\sim$ 0.6 keV and 
$n_et$ $\sim$ 10$^{11}$ cm$^{-3}$ s \citep{hughes98}. The fitted 
electron temperature and the ionization age are consistent with the Sedov 
evolution without significant alteration of the preshock ISM by the stellar 
winds from the progenitor (the model suggested by Shull et al. [1985]), 
which supports the overall picture proposed by Vancura et al. (1992). 
The overall X-ray spectrum was fitted with low metal abundances, 
which is also consistent with the optical/UV spectral analysis. 

The position of the outburst of the soft gamma-ray repeater (SGR) 
SGR 0526$-$66 in 1979 March 5 has been known to be within the 
boundary of N49 \citep{cline82} and is consistent with the unresolved 
X-ray source RX J05260.3$-$660433 \citep{rothschild94}, suggesting 
the identification of this source as the X-ray counterpart of SGR 
0526$-$66. This unresolved X-ray source appears spatially associated 
with SNR N49 based on its positional coincidence within the boundary 
of the SNR although an incidental alignment along the line of sight 
between the two cannot be ruled out \citep{gaensler01}. Although an 
8 s pulsation from SGR 0526$-$66 has been detected during the $\gamma$-ray
burst in 1979 \citep{mazets79}, the X-ray timing/spectral analysis of 
this unresolved X-ray source has been limited with previous detectors 
(e.g., Marsden et al. 1996).
No counterparts in the infrared and radio bands have been found for 
SGR 0526$-$66 \citep{dickel95,smith98,fender98}. These results from 
SGR 0526$-$66 share typical characteristics of the Anomalous X-ray 
Pulsars (AXP), provided that the detection of the X-ray pulsar is 
confirmed. High resolution X-ray imaging with the moderate spectral 
capability of {\it Chandra} observations would be critical to 
confirm the detection of the X-ray counterpart of SGR 0526$-$66. 

We here report the results from the imaging and spatially resolved 
spectral analysis of SNR N49 from the {\it Chandra} observations. 
While this work is primarily devoted to the analysis of the diffuse 
X-ray emission from SNR N49, we also report some preliminary results 
from the detected point source, the X-ray counterpart of SGR 0526$-$66.
We describe the observation and the data reduction in \S \ref{sec:obs}. 
The image and spectral analyses are presented in \S \ref{sec:analysis}, 
and a discussion in \S \ref{sec:disc}. A summary will be presented 
in \S \ref{sec:summary}.

\section{\label{sec:obs} OBSERVATIONS \& DATA REDUCTION}

SNR N49 was observed with the Advanced CCD Imaging Spectrometer 
(ACIS) on board the {\it Chandra X-Ray Observatory} on 2001 
September 15 as part of the {\it Chandra} Guaranteed Time 
Observation (GTO) program (ObsID 1041). The ACIS-S3 chip was chosen to 
utilize the best sensitivity and energy resolution of the 
detector in the soft X-ray bands. The pointing ($\alpha$$_{2000}$ 
= 05$^{h}$ 25$^{m}$ 26$^{s}$.04, $\delta$$_{2000}$ = 
$-$65$^{\circ}$ 59$'$ 06$\farcs$9) was roughly toward the X-ray 
centroid of the nearby LMC SNR N49B with the aim point shifted 
by 245 detector pixels ($\sim$2$'$) from the nominal S3 chip 
aim point in order to simultaneously observe both of N49 and
N49B on the same CCD chip. This pointing direction results in
an angular offset of $\sim$6$\farcm$5 from the aim point 
at the position of N49. The results from the data analysis 
of the pointed target SNR N49B are presented elsewhere 
\citep{park02b}. 

We have utilized the data reduction techniques developed at 
the Pennsylvania State University for correcting the spatial 
and spectral degradation of the ACIS data caused by radiation 
damage, known as charge transfer inefficiency (CTI) 
\citep{townsley00}. The expected effects of the CTI correction
include an increase in the number of detected events and 
improved event energies and energy resolution 
\citep{townsley00,townsley02a}. We screened the data with the 
flight timeline filter and then applied the CTI correction 
before further data screenings by status and grade. We 
removed ``flaring'' pixels and 
selected {\it ASCA} grades (02346). The overall lightcurve 
was examined for possible contamination from time-variable 
background. The overall lightcurve from the whole S3 chip 
shows relatively high background (by a factor of $\ga$2) 
for the last $\sim$10\% of the observation time. This 
contamination from the variable background is however 
negligible for the N49 source region ($\sim$1$\farcm$4 diameter)
and we ignore it in the data analysis. After these data reduction 
process, the effective exposure was 34 ks that contains $\sim$241 
k photons in the 0.3 $-$ 8 keV band for the whole remnant within 
a 1$\farcm$4 diameter circular region.

During {\it Chandra} cycle 1 (2000 January 4), N49 was 
observed with the ACIS-S3 as a Guest Observer (GO) program, 
which is now publicly available in the archive (ObsID 747). 
The pointing was toward the position of SGR 0526$-$66 
($\alpha$$_{2000}$ = 05$^{h}$ 26$^{m}$ 00$^{s}$.7, 
$\delta$$_{2000}$ = $-$66$^{\circ}$ 04$'$ 35$\farcs$0). 
In this observation, instead of the whole CCD and the standard 
3.2 s frame time, a 128-row subarray with a 0.4 s frame time was 
used for the temporal study of the target object SGR 0526$-$66. 
Due to the selection of the subarray of the CCD chip, the coverage 
of the SNR was incomplete and only the northern parts of the SNR 
were observed. The total exposure was $\sim$40 ks. Although the 
angular coverage is incomplete, this on-axis observation provides 
better angular resolutions than our GTO data, thus we utilize these 
archival data as supplements in the analysis. We have reprocessed 
the raw archival data in exactly the same way as the GTO data were 
processed, including the CTI correction.

\section{\label{sec:analysis} ANALYSIS \& RESULTS}

\subsection{\label{subsec:image} X-ray Color Images}

\begin{figure}[]
\figurenum{1}
\centerline{{\includegraphics[angle=0,width=8cm]{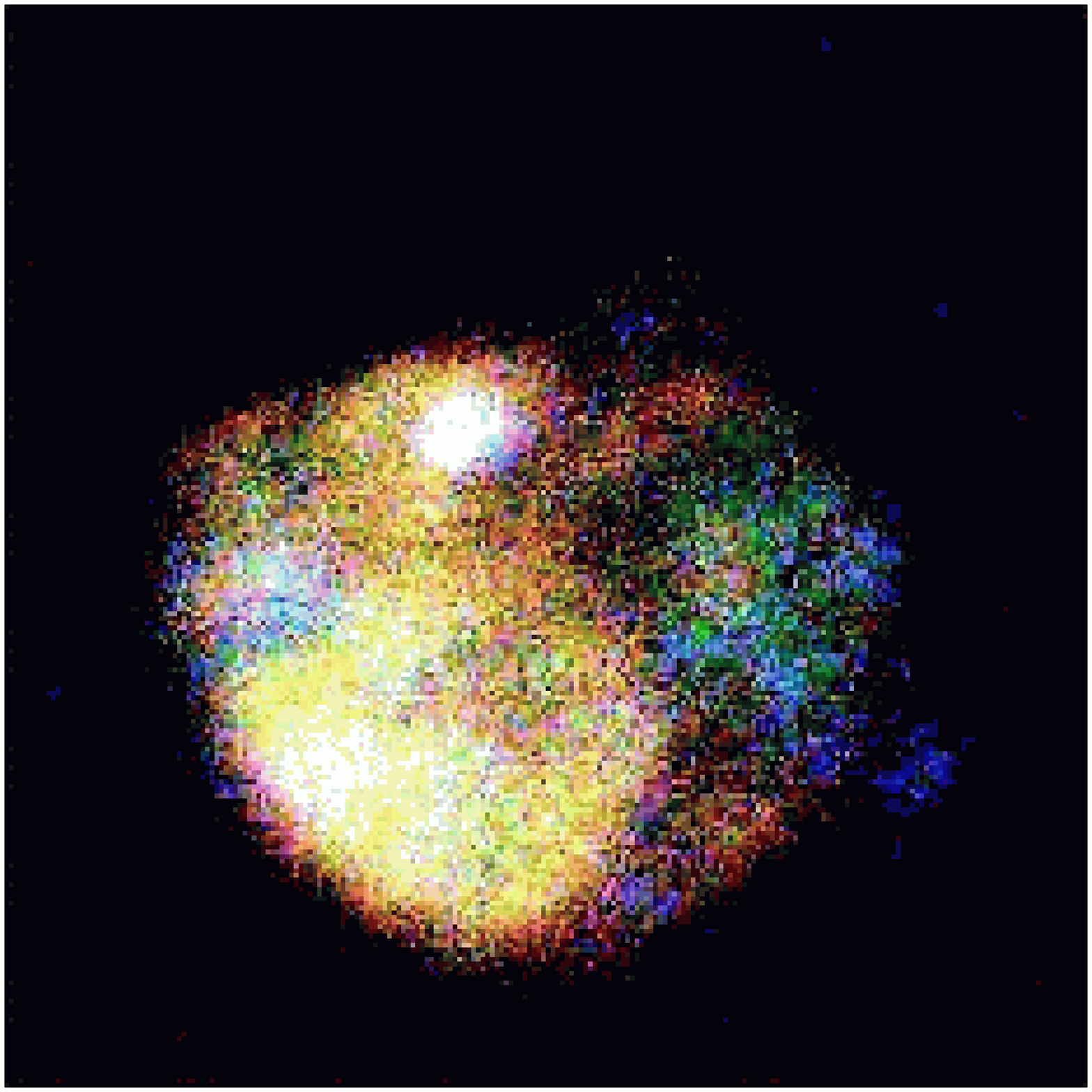}}
{\includegraphics[angle=0,width=7.7cm]{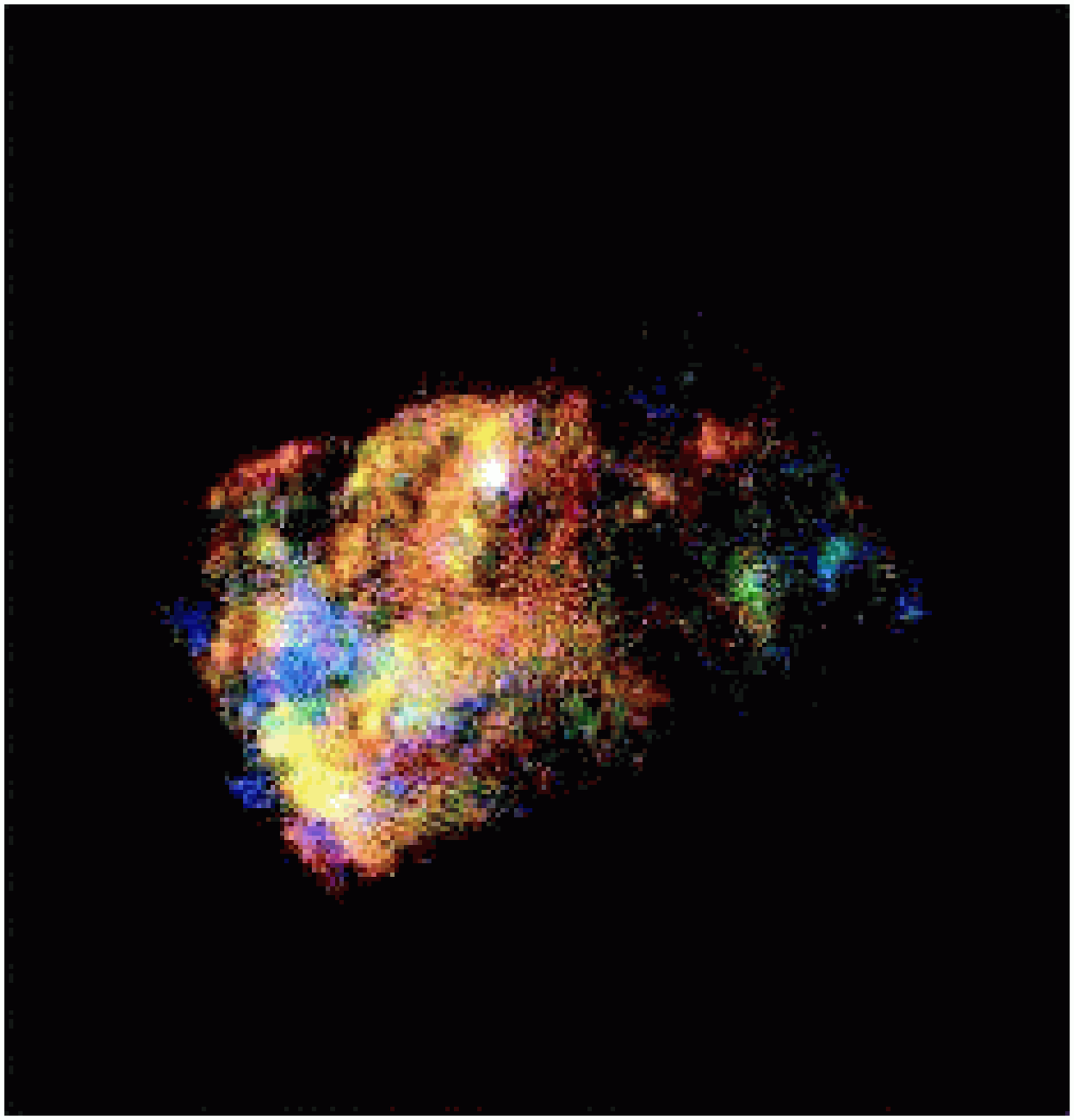}}}
\figcaption[]{The true-color ACIS images of N49. {\it Left:} (a) 
the color image from the GTO data, which is off-axis by 
$\sim$6$\farcm$5. {\it Right:} (b) the color image from the on-axis 
archival data. The pointed target was SGR 0526$-$66. In both images, 
red represents the 0.3 $-$ 0.75 keV band, green is 0.75 $-$ 1.6 keV, 
and blue represents 1.6 $-$ 8.0 keV band photons. 
\label{fig:fig1}}
\end{figure}

The ``true-color'' X-ray images of SNR N49 from the {\it 
Chandra}/ACIS observations are presented in Figure \ref{fig:fig1}.
Figure \ref{fig:fig1}a is the image from the GTO observation
and Figure \ref{fig:fig1}b is the one from the archival GO data. The 
differences in the point spread function (PSF) between two images 
due to the angular offset of $\sim$6$\farcm$5 from the aim point in 
the GTO data are apparent. The high resolution on-axis archival 
data reveal the detailed diffuse filamentary structure, particularly 
in the east of the SNR, far better than the off-axis GTO data. 
The detection of the point source as the counterpart of SGR 0526$-$66 
is evident in the middle of the northern part of the SNR as shown by 
a bright white spot (Figure \ref{fig:fig1}b). This point source 
is also detected in the off-axis image (Figure \ref{fig:fig1}a) 
but appears extended because of the PSF degradation. Despite the
superior angular resolution of the on-axis data, the selection
of the subarray of the CCD allows only a partial coverage of 
the SNR: i.e., about a third of the SNR in the south-west was 
not covered (Figure \ref{fig:fig1}b). The imaging and spectral 
analysis over the whole remnant is important to understand the global
nature of the SNR and we primarily utilize our GTO data for the data 
analysis. In order to take advantage of the high angular resolution, 
we use the archival data for supplementary purposes.

\begin{figure}[]
\figurenum{2}
\centerline{\includegraphics[angle=-90,width=15cm]{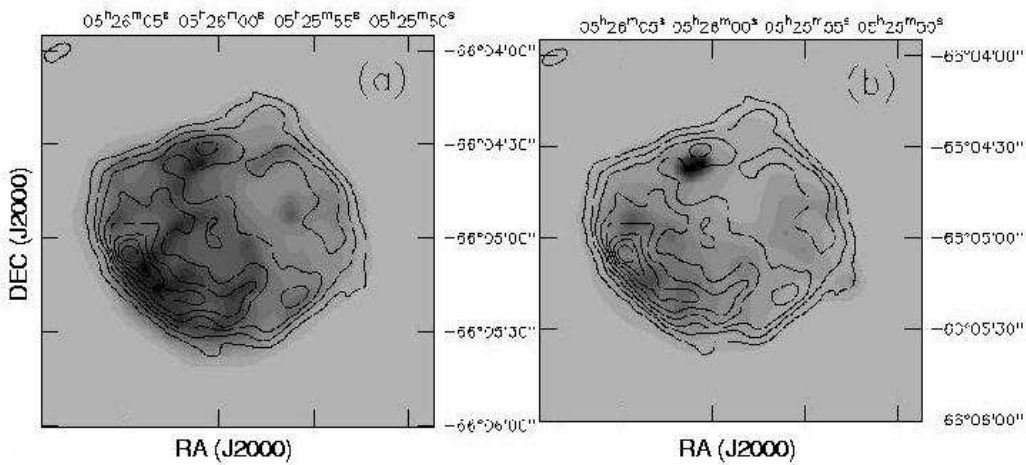}}
\figcaption[]{ {\it Left:} (a) The soft X-ray (0.3 $-$ 0.75 keV) image
of N49 overlaid with the 6 cm radio contours. {\it Right:} (b)
The hard X-ray (1.6 $-$ 8.0 keV) image of N49 overlaid with the 6 cm
radio contours. In both images, the X-ray images have been adaptively
smoothed and the darker gray-scales represent higher X-ray intensities. 
The radio data are from the 6 cm continuum observed with Australia 
Telescope Compact Array (ATCA) (provided by John Dickel). The ATCA 
data have an 1$\farcs$58 $\times$ 1$\farcs$44 elliptical half-power 
beamwidth.
\label{fig:fig2}}
\end{figure}

In Figure \ref{fig:fig1}, the soft X-ray emission (red-yellow)
reveals various filamentary structures across the SNR. The 
circumferential shell-like soft  X-ray features around the SNR are 
spatially coincident with the radio shells (Figure \ref{fig:fig2}a), 
representing the blast wave shock front. By comparison, the hard X-ray 
emission features show little correlation with the radio intensities 
(Figure \ref{fig:fig2}b). Based on astrophysical considerations, 
we believe that it is reasonable to suppose that shocks are reflecting 
off the denser molecular gas in the east and may be propagating back 
into the interior of N49. This could be a partial explanation for 
the higher temperatures we find in the spectral analysis of the 
eastern region (\S \ref{subsubsec:east}), although a definitive 
conclusion would require a detailed investigation of the hydrodynamics 
of N49 that is beyond the scope of this work. Other evidence, albeit 
more speculative, in support of reflected shocks in the eastern region 
of N49 include (1) the morphology of the hard X-rays which peak closer 
to the interior on the eastern side than the soft X-rays do (Figure 
\ref{fig:fig2}), and (2) the clear arc-like feature in the Si EW map 
(\S \ref{subsec:ew} and Figure \ref{fig:fig5}f) which traces hot gas, 
that perhaps delineates the edge of the inward moving reflected shock. 
The western half is, for the most parts, dominated by hard X-ray 
emission (green-blue), which may indicate that the blast wave is 
propagating through a low density ISM, as suggested by radio data 
\citep{dickel98}. 

\begin{figure}[]
\figurenum{3}
\centerline{\includegraphics[angle=0,width=7.5cm]{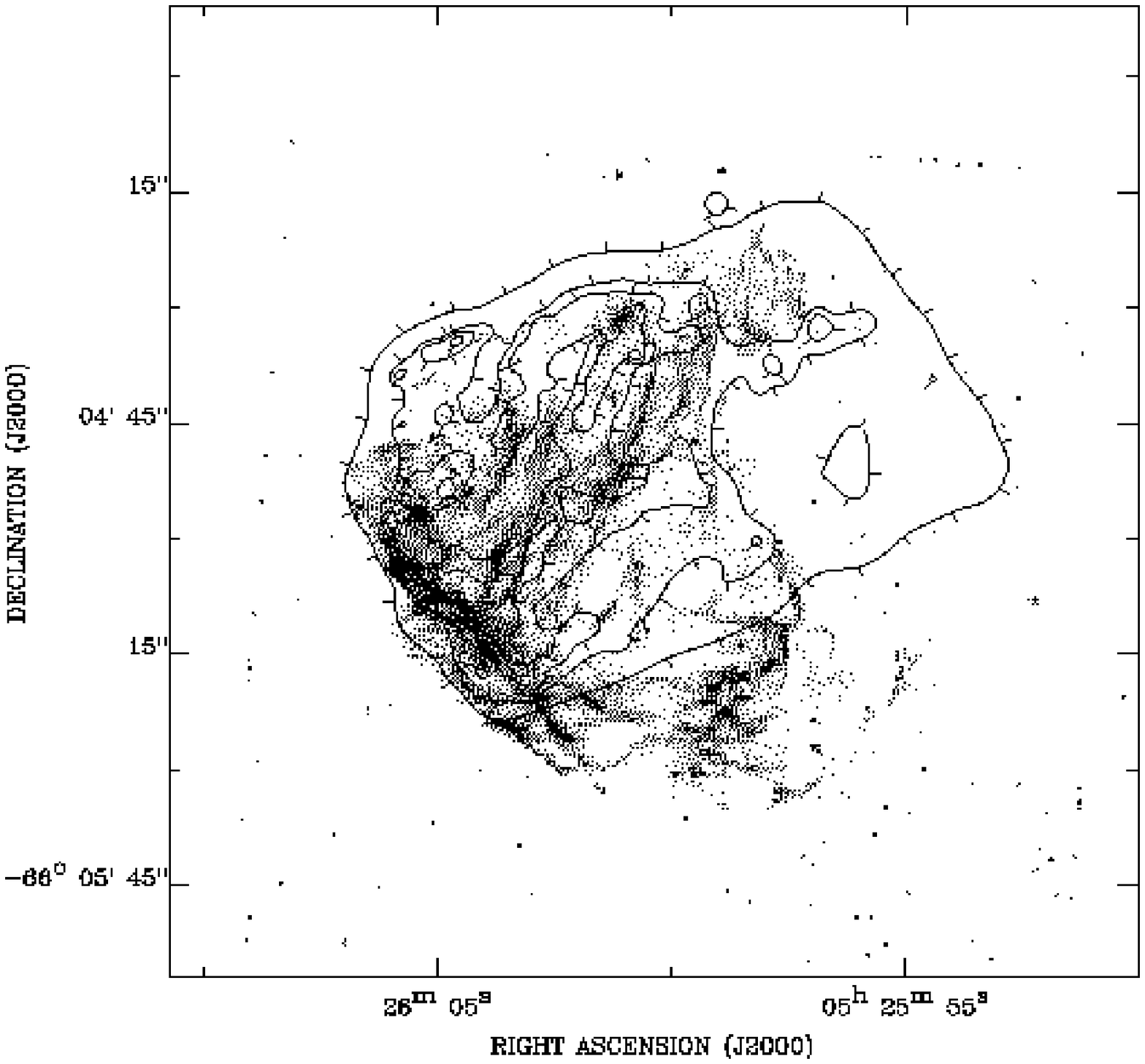}
\includegraphics[angle=0,width=7.5cm]{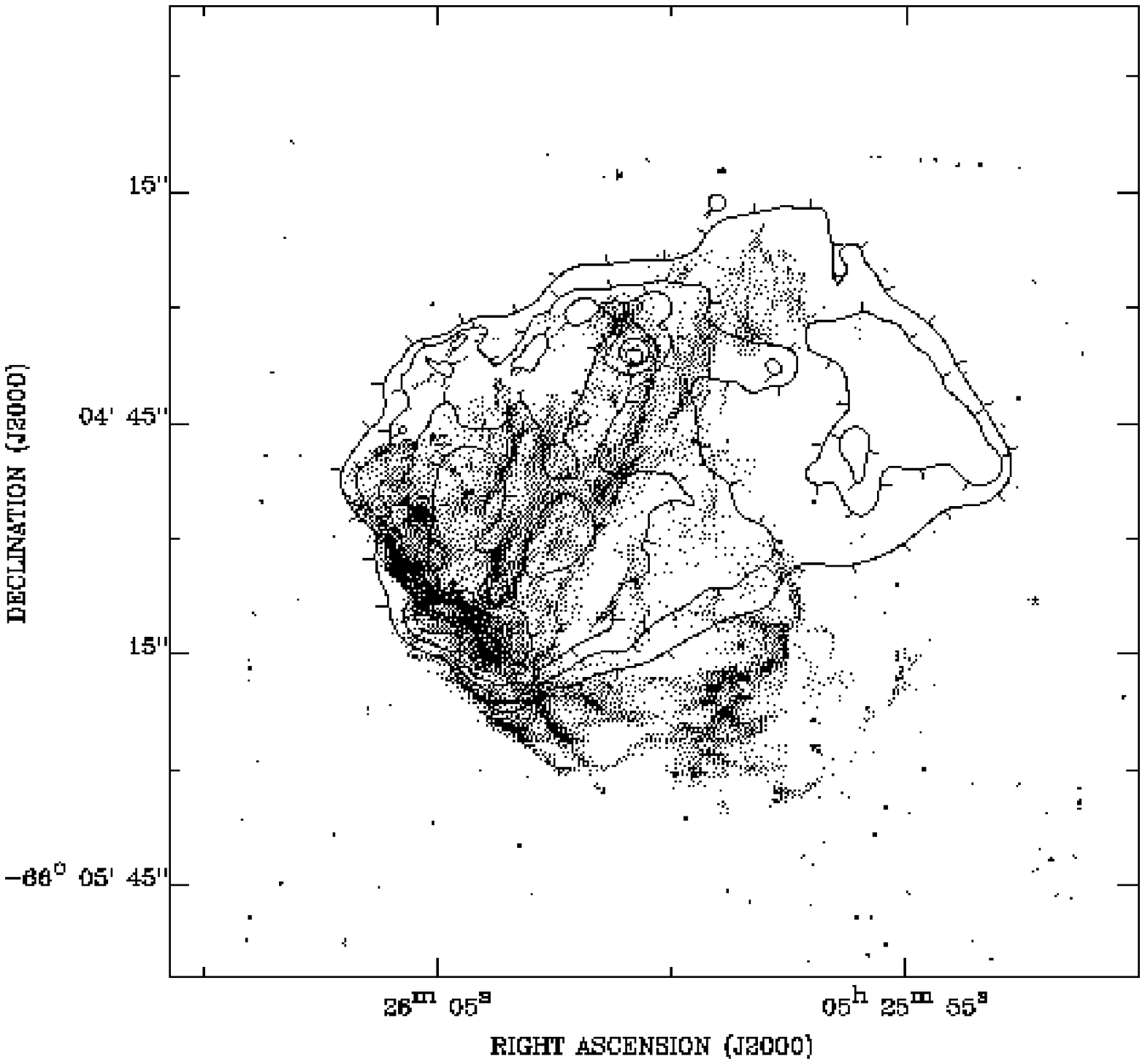}}
\figcaption[]{ {\it Left:} (a) The optical O{\small III} image of N49 
overlaid with the soft X-ray (0.3 $-$ 0.75 keV) contours from the archival 
GO data. (b) The optical O{\small III} image of N49 overlaid with the hard
X-ray (1.6 $-$ 8.0 keV) contours from the archival GO data. Darker
gray-scales indicate higher optical intensities. The X-ray images have been
adaptively smoothed and the ``downstream'' tickmarks indicate the direction 
of decreasing X-ray count rates. Optical image is an HST WFPC2 image taken 
through the F5002 narrow-band filter. Cosmic rays have been removed and 
WFPC chips mosaicked for this image.
\label{fig:fig3}}
\end{figure}

The brightest diffuse emission is in the southeast, as recognized 
by previous X-ray observations, where the SNR is interacting with 
dense molecular clouds. The overall bright X-ray emission in the 
eastern half of the SNR is significant in all colors and likely 
has complex spectral structure. While filamentary soft, red-yellow 
emission prevails in the eastern half of the SNR, there are apparent 
blue regions in the east (primarily to the immediate north of the 
southeast brightest region; Figure \ref{fig:fig1}). This diffuse X-ray 
emission with spectral variations appears localized to produce 
spatially distinctive layers of X-ray colors in the eastern half of the 
SNR. The interactions of the blast wave shock with the clumpy dense 
molecular clouds in these regions appear responsible for the observed 
spatio-spectral substructures of the diffuse X-ray emission (see \S 
\ref{subsec:spec}). For example, the bright filametary emission 
features from the optical O{\small III} images are largely anti-correlated 
with the blue, hard X-ray intensity and more correlated with the soft, 
red X-ray emission (Figure \ref{fig:fig3}a and Figure \ref{fig:fig3}b). 
The optical emission from N49 is believed to originate primarily from slow 
radiative shocks entering dense molecular clouds, while the X-ray 
emission is in general from fast shocks propagating through the 
intercloud medium \citep{vancura92}. The high resolution X-ray images 
from the ACIS data reveal that the hard X-rays are indeed from 
intercloud regions where little optical emission originates, and that the 
soft X-ray emission is spatially more coincident with the bright 
optical emitting regions. These features support the interactions of 
the blast wave with the clouds as the origin of the X-ray spectral 
variations observed in the color images.

We detect an interesting, previously unknown feature of N49 with 
our {\it Chandra} GTO data: i.e., the blue, hard X-ray knot in the 
southwest, beyond the main boundary of the SNR (Figure 
\ref{fig:fig1}a), from which faint radio emission is also associated
(Figure \ref{fig:fig2}b). The extended shape of this feature appears 
caused by the off-axis PSF and the true morphological nature of this 
emission is uncertain with the current off-axis data. We cannot 
utilize the high resolution archival data, because this feature lies
outside the field of view (Figure \ref{fig:fig1}b). We find no proper 
counterparts for this feature in other wavelengths, which suggests 
that it is unlikely a projection of a background/foreground point 
source (\S \ref{subsubsec:bullet}). The X-ray spectrum of this feature 
indicates metal-rich thermal emission with strong atomic emission 
lines rather than a power law dominated spectrum (\S 
\ref{subsubsec:bullet}). These spectral features support the proposition 
that this hard X-ray knot is a part of the SNR, probably fragment(s) 
of SN ejecta. 

\subsection{\label{subsec:ew} Equivalent Width Images}

\begin{figure}[]
\figurenum{4}
\centerline{\includegraphics[angle=-90,width=12cm]{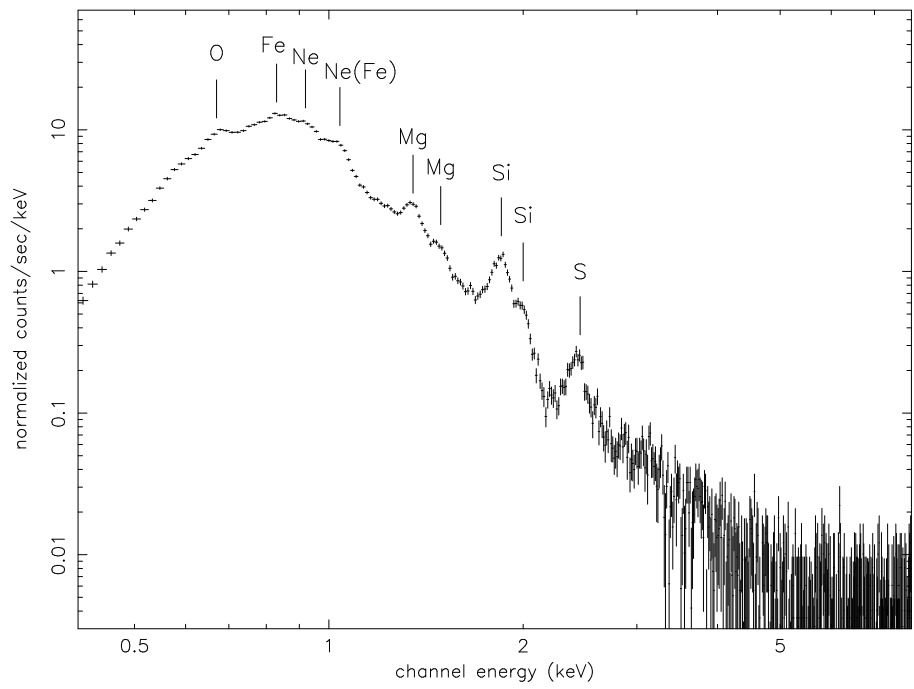}}
\figcaption[]{The overall spectrum of N49.
\label{fig:fig4}}
\end{figure}

\begin{figure}[]
\figurenum{5}
\centerline{\includegraphics[angle=0,width=15cm]{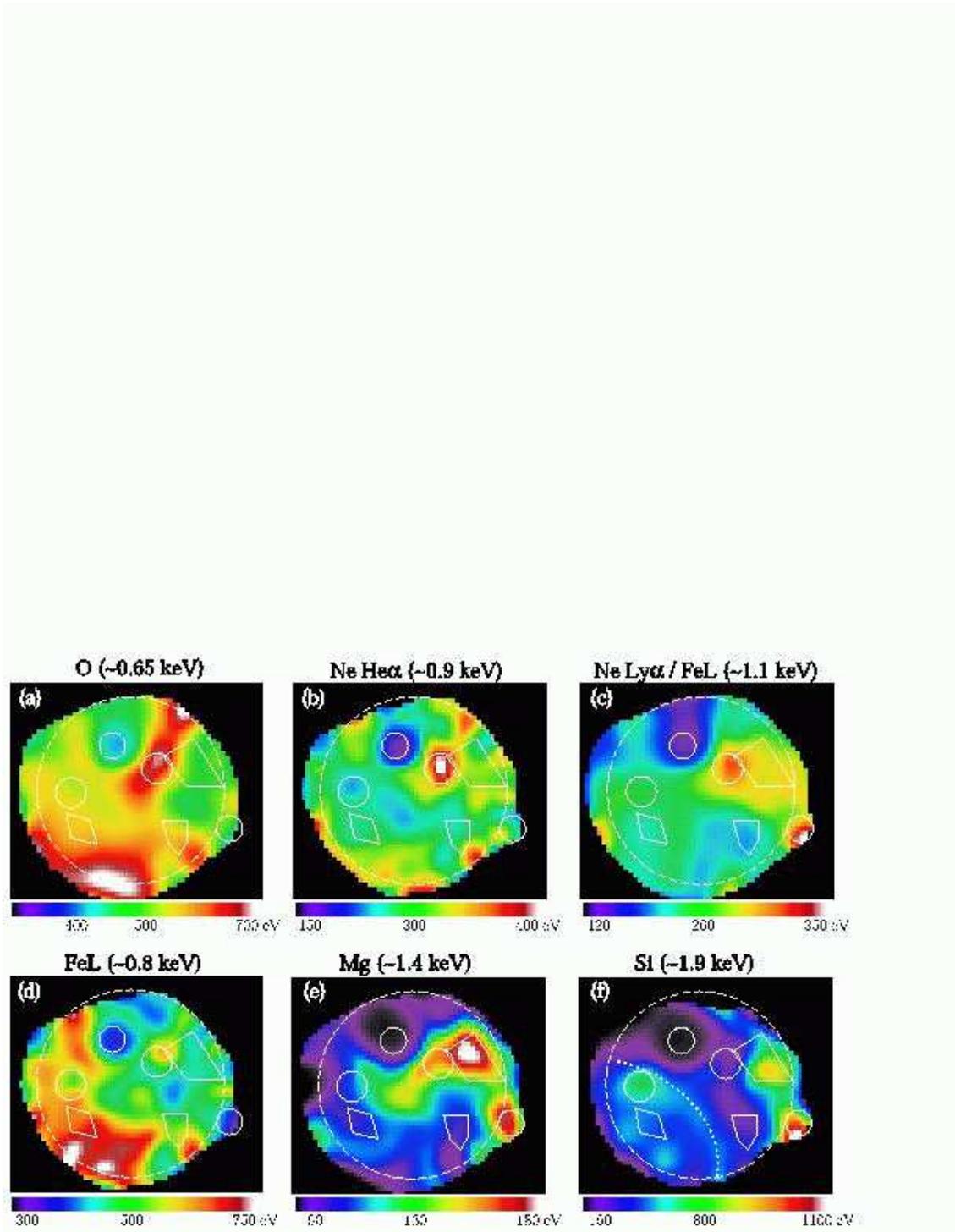}}
\figcaption[]{The EW images of N49. In order to remove the noise 
due to the background, EWs are set to zero where the estimated 
continuum fluxes are low ($<$5\% of the maximum).
The small regions where the actual spectrum was extracted for the
spectral analysis, as also presented in Figure \ref{fig:fig6}, are
marked. The large dashed circle is to represent the main boundary
of the SNR as is presented in Figure \ref{fig:fig6}. In (f),
the dotted curve is to present the arc-like feature of the Si EW.
\label{fig:fig5}}
\end{figure}

The overall spectrum of N49 shows some broad atomic emission 
lines (Figure \ref{fig:fig4}) and we generate {\it equivalent 
width} (EW) images of K-shell line emission from the atomic 
elements O, Ne, Mg, Si, S, and the L-shell line emission of Fe 
by selecting photons around the broad line complexes 
(Table \ref{tbl:tab1}), following the methods described by 
Park et al. (2002a). Images in the bandpasses of interest for 
both the lines and continuum were extracted with 2$\arcsec$ 
pixels and smoothed by a Gaussian with $\sigma$ = 7$\arcsec$. 
The underlying continuum was calculated by logarithmically 
interpolating between images made from the higher and lower 
energy ``shoulders'' of each broad line. The estimated 
continuum flux was integrated over the selected line width and 
subtracted from the line emission. The continuum-subtracted line
intensity was then divided by the estimated continuum on a 
pixel-by-pixel basis to generate the EW images for each element. 
In order to avoid noise in the EW maps caused by poor photon 
statistics near the edge of the SNR, we have set the EW values 
to zero where the estimated continuum flux is low. We also set 
the EW to zero where the integrated continuum flux is greater 
than the line flux. We present the EW images for O, Ne, Mg, Si, 
and Fe (Figure \ref{fig:fig5}). The S EW image is unreliable due 
to the low photon statistics. The broad line complexes at $\sim$0.8 
$-$ 1.1 keV are from various ionization states of Ne K and Fe L-shell 
line emission, and we present tentative identifications in Figure 
\ref{fig:fig5} and Table \ref{tbl:tab1}. Although Mg and Si lines 
show He-like and H-like substructures in the overall spectrum 
(Figure \ref{fig:fig4}), we present the EW images from the combined 
broad lines in Figure \ref{fig:fig5} since the H-like lines of these 
elements are too weak to separately produce useful EW images over 
the whole SNR. 

The O EW distribution is generally consistent with the optical 
O{\small I,III} intensity distributions and shows higher EW in 
the east and south of the SNR (Figure \ref{fig:fig5}a; Vancura
et al. 1992). The Fe EW (Figure \ref{fig:fig5}d) shows similar 
distribution to the O map. Ne/Fe and Mg EW (Figure \ref{fig:fig5}b, 
Figure \ref{fig:fig5}c, and Figure \ref{fig:fig5}e) distributions 
appear to be opposite to O, showing higher intensities in the west 
of the SNR. The Si EW is relatively strong in both east and west
(Figure \ref{fig:fig5}f). The O and Fe EW enhancements are spatially 
correlated with the location of the dense molecular clouds where 
the X-ray and optical emission is generally bright (Figure 
\ref{fig:fig5}a and Figure \ref{fig:fig5}d). Since the X-ray 
emission in these regions is presumably dominated by the shocked 
dense ISM, these high O and Fe EWs are unlikely due to abundance 
enhancements, but appear to trace shock temperature variations. 
Optical spectroscopic studies have indicated that metal abundances 
of N49 are close to the LMC values \citep{russell90,vancura92} and 
support temperature variations, rather than abundances, 
as the origin of these EW distributions. Based on the plane-parallel
shock model with a moderate ionization timescale ($n_et$ $\sim$
10$^{12}$ cm$^{-3}$ s) and the LMC abundances, the Fe L EW (Figure 
\ref{fig:fig5}d) is in fact representative of the broad line complex 
at $E$ $\sim$ 0.8 keV, which is characteristic of a ``cool'' plasma 
($kT$ $\la$ 0.5 keV) as opposed to the ``hot'' component ($kT$ 
$\sim$ 1 keV) that would produce a prominent Fe L line complex at 
$E$ $\sim$1 keV. 

In direct contrast, the H-like Ne (and/or hot Fe L) EW (Figure 
\ref{fig:fig5}c), which is sensitive to high temperatures of $kT$
$\sim$ 1 keV, is enhanced in the west where the blast 
wave is expanding into a low density ISM. These EW distributions 
may thus indicate the presence of an {\it undisturbed} (by dense ISM) 
blast wave shock in the west, which may have a higher temperature 
than that in the eastern side of the SNR. If the dominant contributor 
to the enhanced EW at $E$ $\sim$ 1.1 keV is Ne, comparisons between 
He-like and H-like Ne EW images (Figure \ref{fig:fig5}b and Figure 
\ref{fig:fig5}c) may suggest progressive ionizations of Ne
by the reverse shock as detected in other SNRs 
\citep{gaetz00,flanagan01,park02a}. The Si EW enhancements in both 
east and west are intriguing and may suggest the presence of hot 
phases in addition to cool phase shocks in the south-east parts of 
the SNR (see more discussion in \S \ref{subsec:spec}). The Si EW 
enhancements in the southeastern quadrant are largely shell-like, 
as indicated by a dotted line in Figure \ref{fig:fig5}f. This 
may also be interpreted as hot, highly ionized plasma in the 
southeastern regions of the SNR.

The hard X-ray knot in the southwest beyond the SNR boundary, as 
revealed with the true-color image, shows strong enhancements in 
the Si, Mg, and hot Ne/Fe EW images. The actual spectral analysis
indicates that these high EWs are caused by metal abundances as 
well as advanced ionizations and high temperature (see \S 
\ref{subsubsec:bullet}). The enhanced EWs for the elemental species 
caused by high metal abundances support that this hard X-ray knot 
may not be a background extragalactic source, and that it is likely 
emission from metal-rich SN ejecta associated with N49. The EW 
around the detected point source is, on the other hand, very low 
for all elemental species. These low EWs are strong indications of 
the continuum-dominated X-ray spectrum for the point source. This 
is in good agreements with the presumed identification of the point 
source as the neutron star, or the X-ray counterpart of SGR 0526$-$66 
in the quiescent state. 

\subsection{\label{subsec:spec} X-ray Spectra} 

The high resolution ACIS data, for the first time, allow us to perform
spatially-resolved spectral analysis in X-rays for the study of SNR 
N49. Despite the PSF degradations due to the large angular 
offset from the aim point, the GTO data provide moderate angular 
resolution of a few arcsec, which are still sufficient to 
resolve major diffuse structure of N49. Moreover, the complete 
spatial coverage of the SNR with the GTO data is advantageous over 
the archival data for the efficient and consistent study of the 
global structures of the SNR. Combining two data sets may not be 
desirable because of the significant difference in the PSF. We 
thus make a simple and secure approach for the spatially-resolved 
spectral analysis by utilizing the GTO data as the primary data 
and the archival data as the supplementary data: i.e., we perform 
all spectral analysis with the GTO data and then verify the results 
with the archival data as necessary. We extract the spectrum from 
several small regions over the SNR. The regional selections were 
based on the characteristics revealed by the broad band color 
images and the EW images, and are presented in Figure \ref{fig:fig6}. 
In all spectral fittings, we made a low energy cut at the photon 
energy of $E$ = 0.5 keV because of unreliable calibrations 
of the ACIS spectrum in the soft bands. In most cases, we bin the 
data to contain a minimum of 20 counts for the spectral fitting. 
For the spectral analysis of our CTI corrected data, we have 
utilized the response matrices appropriate for the spectral 
redistribution of the CCD, as generated by Townsley et al. (2002b).
The low energy ($E$ $<$ 1 keV) quantum efficiency (QE) of the ACIS
has been degraded because of the molecular contamination on the 
optical blocking filter. We have corrected this time-dependent
QE degradation by modifying the ancillary response function (ARF)
for each extracted spectrum by utilizing the IDL version of
ACISABS software developed by Chartas et al. (2002).

\subsubsection{\label{subsubsec:west} SNR West: Single-Phase Shock}

\begin{figure}[]
\figurenum{6}
\centerline{\includegraphics[angle=0,width=10cm]{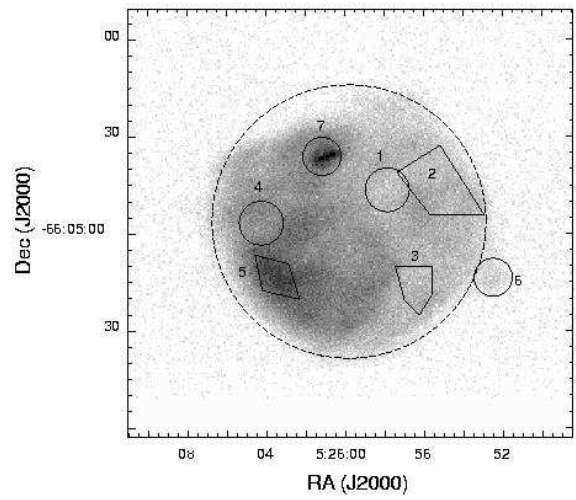}}
\figcaption[]{Gray-scale broad band (0.3 $-$ 8.0 keV) image of N49. 
Darker gray-scales indicate higher X-ray internsities.
The regions used for the spectral analysis are marked. The large dashed 
circle is to represent the main boundary of the SNR as is marked in 
Figure \ref{fig:fig5}.
\label{fig:fig6}}
\end{figure}

\begin{figure}[]
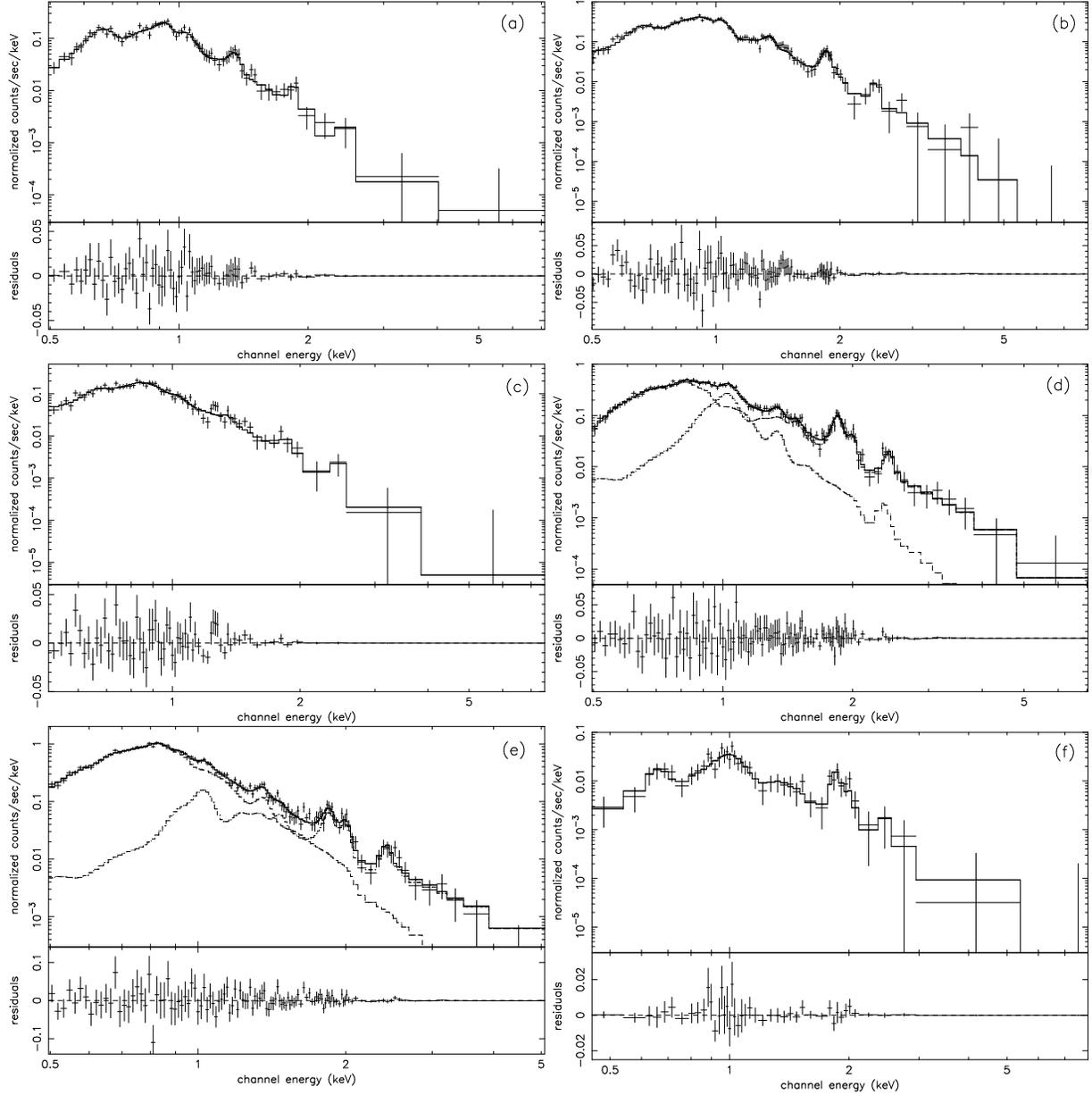

\figurenum{7}
\centerline{\includegraphics[angle=-90,width=8cm]{fig7a.ps}
\includegraphics[angle=-90,width=8cm]{fig7b.ps}}
\centerline{\includegraphics[angle=-90,width=8cm]{fig7c.ps}
\includegraphics[angle=-90,width=8cm]{fig7d.ps}}
\centerline{\includegraphics[angle=-90,width=8cm]{fig7e.ps}
\includegraphics[angle=-90,width=8cm]{fig7f.ps}}
\figcaption[]{Regional spectra of N49 as extracted from regions
marked in Figure \ref{fig:fig6}. (a) $-$ (f) represents region 
1 $-$ 6, respectively. For regions 4 and 5, we have fitted 
the spectrum with two-temperature model and each component is 
displayed with dashed lines in (d) and (e).
\label{fig:fig7}}
\end{figure}

We select three regions from the western half of the SNR (regions
1, 2, and 3; Figure \ref{fig:fig6}). These regions are where the 
broad band surface brightness is relatively low, while the EW 
images reveal some noticeable features (Figure \ref{fig:fig5} and 
Figure \ref{fig:fig6}). Region 1 is where O and Ne EWs are enhanced 
and contains $\sim$3500 photons. Region 2 is where Ne/Fe, Mg and Si 
are enhanced containing $\sim$8100 counts. Region 3 shows low EW for 
most of the elements and contains $\sim$3200 photons. We fit these 
spectra with the plane-parallel shock model \citep{borkowski01} 
as presented in Figure \ref{fig:fig7}a, Figure \ref{fig:fig7}b, and
Figure \ref{fig:fig7}c. The elemental abundances were fixed for 
He (= 0.89), C (= 0.30), N (= 0.12), Ca (= 0.34) and Ni (= 0.62) 
at values for the LMC \citep{russell92} (hereafter, all abundances 
are with respect to the solar; Anders \& Grevesse 1989) since the 
contribution from these species in the spectral fitting is expected 
to be insignificant in the selected energy range. Other elements 
were allowed to vary. The results of the spectral fittings for the 
west regions of the SNR are summarized in Table \ref{tbl:tab2} and 
Table \ref{tbl:tab3}.

These spectra in the western half of the SNR can be described 
by a thermal plasma with the best-fit $kT$ of 0.52 $-$ 0.65 keV. 
The implied absorptions are $\sim$0.6 $-$2.7 $\times$ 10$^{21}$ 
cm$^{-2}$. These results are consistent with the results from
the {\it ASCA} data \citep{hughes98}. 
The fitted abundances are generally 
consistent with the EW distributions. The results from the region 2 
spectrum are perhaps the most interesting. This is where Ne/Fe, Mg, and
Si EWs are enhanced and the best-fit abundances for those elemental
species (except for Fe) are also relatively high, although the 
magnitudes of the enhancements are moderate only at around the 
solar level. The fitted abundances for this region suggest that 
the high EW responsible for the emission line at $E$ $\sim$ 1.1 keV 
(Figure \ref{fig:fig5}c) is due to highly ionized H-like Ne rather 
than hot Fe. The best-fit shock temperature and the ionization 
timescale for this region are higher than region 1 in which He-like
Ne EW is enhanced (Figure \ref{fig:fig5}b). This supports the 
advanced ionization in region 2 as suggested by the EW images. 
The region 3 spectrum is on the other 
hand fitted with low LMC-like abundances for all elements. The broad 
band X-ray surface brightness is also low for this region, indicating 
X-ray emission from the shocked low density LMC ISM. 

\subsubsection{\label{subsubsec:east} SNR East: Multi-Phase Shock}

We select two regions from the eastern half of the SNR (regions
4 and 5; Figure \ref{fig:fig6}). Region 4 is within the bright
eastern rim which is evidently dominated by the hard X-ray
emission as displayed in Figure \ref{fig:fig1}. A total of
$\sim$10000 photons are extracted from region 4. Region 5 is the 
brightest knot in the southeast, where we extract $\sim$17000 counts. 
We fit these spectra in the same way as we did for the western 
regions: i.e., we use the plane-parallel shock model with O, Ne,
Mg, Si, S, and Fe abundances variable in the fittings. Both of the 
spectra in the eastern half of the SNR show significant Si 
Ly$\alpha$ line ($E$ $\sim$ 2.0 keV) as well as strong soft X-ray 
emission ($E <$ 1 keV) (Figure \ref{fig:fig7}d and Figure 
\ref{fig:fig7}e). In order to adequately describe these spectral 
features, a hot component shock with $kT$ $\ga$ 1 keV is required 
in addition to the soft component with $kT$ $\sim$ 0.4 $-$ 0.5 keV. 
The hot components have larger ionization timescales ($n_et$ $\ga$ 
10$^{12}$ cm$^{-3}$ s) than the cool components ($n_et$ $\sim$ 
10$^{11}$ cm$^{-3}$ s). The fitted ionization timescales for the 
hot component are not constrained in the upper bounds and a thermal 
plasma in the complete collisional ionization equilibrium (CIE) 
may also fit the data with the equivalent statistical significance. 
The implied absorbing columns are $N_H$ $\sim$ 3 $-$ 4 $\times$ 
10$^{21}$ cm$^{-2}$ for these regions.

The soft component model for the region 5 spectrum indicates low metal 
abundances of $\la$ 0.5 solar for fitted elemental species. This
soft component dominates the X-ray emission at $E$ $\la$ 1 keV, 
which consistently supports that the enhanced O and Fe EWs (Figure 
\ref{fig:fig5}a and Figure \ref{fig:fig5}d) are not caused by high 
abundances. The best-fit model for the hard component implies 
overabundances (typically $\sim$ 1 $-$ 2 solar) compared with the LMC 
abundances for some elemental species. The statistical uncertainties for 
the fitted abundances are however large enough that the 
reality of these high abundances are inconclusive with the current 
data. For example, this region 5 spectrum may also be fitted with 
subsolar abundances, for both of the soft and hard components of 
the shock, with equivalent statistical significance based on the F-test. 
The EWs distributions in the east of the SNR are therefore more likely 
dominated by the shock temperature and the ionization states rather 
than by metal abundances. The fitted  model parameters for the region 4 
spectrum are also generally consistent with the results from region 
5, which also supports the temperature and ionization states for 
the origins of the observed EW distributions. We note that region 4 
EWs appear relatively weak for the atomic lines at low photon energies 
(O and Fe; Figure \ref{fig:fig5}a and Figure \ref{fig:fig5}d) and 
stronger in higher energies (Si; Figure \ref{fig:fig5}f) compared 
with region 5, even though there is no significant difference in 
the fitted shock parameters between these two regions. These features 
may imply a difference in the local interstellar environments due 
to the clumpy structure of the interacting dense molecular clouds: 
i.e., a density gradient of the clouds between these two regions 
might have caused deficient soft X-ray emission in region 4 as 
displayed by the lack of red, soft emission there (Figure 
\ref{fig:fig1}). The anticorrelations between optical and hard X-ray 
emission are supportive of this case. The fitted volume emission 
measure ($EM$) of the soft component plasma for the region 5 spectrum is 
indeed significantly higher than that for region 4, which also supports 
this interpretation. The results of the spectral fittings for the 
east regions of the SNR are summarized in Table \ref{tbl:tab4}.

\subsubsection{\label{subsubsec:bullet} Extended Feature Beyond the
Shock Boundary}

The hard X-ray knot in the southwest beyond the main boundary of the 
SNR (region 6 in Figure \ref{fig:fig6}) contains $\sim$800 counts 
and we binned the spectrum to contain a minimum of 12 counts per bin.
The spectrum can be fitted with a single-temperature plane-parallel
shock (Figure \ref{fig:fig7}f). The spectrum features strong atomic 
emission lines and is most likely emission from metal-rich ejecta. 
The strong Si Ly$\alpha$ line ($E$ $\sim$ 2.0 keV) constrains the 
plasma temperature to $kT$ $\ga$ 1 keV. Although statistical
uncertainties are large due to the poor photon statistics of this
faint feature, the best-fit model indicates strong enhancements 
in metal abundances at several times larger than solar abundances. 
The broad band line features between $\sim$0.8 keV and $\sim$1.2 
keV represent Fe L line complex as well as He-like and/or H-like 
Ne K lines. With the photon statistics of the current data, it
is difficult to identify these individual atomic lines. We have 
thus attempted fitting the spectrum by assuming dominant contributions 
from Ne or Fe as well as from both elements. The overall fits are
statistically indistinguishable among these cases. We however
note that the model fits Fe as dominant contributor (Fe abundance
$\sim$ 0.8, Ne $\sim$ 0) when both Ne and Fe are simultaneously 
allowed to vary. Even with Fe abundance fixed at LMC value, the 
fitted Ne abundance is still low ($\sim$0.4). We therefore obtain 
the best-fit model by allowing Fe abundance to vary with Ne fixed 
at the LMC abundance. The fitted parameters are summarized in 
Table \ref{tbl:tab2} and Table \ref{tbl:tab3}.

Although the metal-rich spectrum indicates that this feature is
likely emission from shocked SN ejecta, the isolated morphology 
of this hard X-ray knot outside of the main boundary of the SNR 
may suggest the possibility of a projected background object. 
The source position for the knot is $\alpha$$_{2000}$ = 05$^{h}$ 
25$^{m}$ 52$^{s}$.53, $\delta$$_{2000}$ = $-$66$^{\circ}$ 05$'$ 
13$\farcs$5, and the ACIS astrometric uncertainties at $\sim$6$'$ 
off-axis is generally $\la$2$\arcsec$ (e.g., Feigelson et al. 2002). 
The {\it Chandra}/ACIS position of SGR 0526$-$66 based on the on-axis 
data has a conservative astrometric uncertainty of $\la$2$\farcs$3
(Kaplan et al. 2001 and references therein). The position of the SGR
of our off-axis data as determined by {\it wavdetect} algorithm is
$\sim$1$\farcs$8 offset from the on-axis SGR position. The source
positions with our off-axis data therefore appear to have
$<$4$\arcsec$ accuracy as a conservative limit. We have searched for 
counterparts around the source position in SIMBAD and NED catalogs as 
well as $\sim$100 multi-wavelengh on-line catalogs available through 
HEASARC. We find no candidate for an extragalactic counterpart within 
1$'$ radius of the source position. The derived X-ray flux is 
$f_{X(0.5-2 keV)}$ $\sim$ 6.0 $\times$ 10$^{-14}$ ergs s$^{-1}$ 
cm$^{-2}$ and $f_{X(2-10 keV)}$ $\sim$ 1.5 $\times$ 10$^{-14}$ ergs 
s$^{-1}$ cm$^{-2}$. Based on the logN $-$ logS relations of the {\it 
Chandra} Deep Field \citep{brandt01}, these X-ray fluxes imply the 
probability of only $\la$10$^{-4}$ within a generous area of 5$\arcsec$ 
radius around the source position for a coincidental detection of an 
extragalactic object. 

We find an LMC star, MACS J0525$-$660\#037 \citep{tucholke96} which 
has an angular offset from the position of the hard X-ray knot of
$\sim$6$\farcs$2. Considering the positional accuracy of the MACS 
catalog ($\sim$0$\farcs$5; Tucholke et al. 1996), the apparent angular 
offset between the hard X-ray knot and optical position of the MACS 
star is significant. We find another nearby LMC star within the 
OB association LH 53 as cataloged (ID \# 1630) by Hill et al. (1995). 
The relatively small angular offset ($\sim$2$\arcsec$) of this star 
from the X-ray hard knot position suggests that this star could be the 
optical counterpart of the X-ray knot. The derived X-ray luminosity
of $L_X$ $\sim$ 3.4 $\times$ 10$^{34}$ ergs s$^{-1}$ (at the distance 
of 50 kpc) for the X-ray knot is however higher than that of typical 
late-type young stellar objects ($L_X$ $\sim$ 10$^{28}$ $-$ 10$^{30}$
ergs s$^{-1}$; e.g., Imanish et al. 2001) by a few orders of magnitude. 
A massive early-type star should be brighter ($L_X$ $\sim$ 10$^{32}$ 
ergs s$^{-1}$; e.g., Seward \& Chlebowski 1982), but may also not 
match the derived high $L_X$ of the hard X-ray knot. Moreover, although 
there are $\sim$80 LH 53 stars within 1$'$ radius of this hard X-ray 
knot position \citep{hill95}, none of them is detected with our
ACIS observation at the flux level of this hard X-ray knot. It is 
unlikely for only one star to be detected with the ACIS at this 
particular position at this flux level.

If this object is a foreground young late-type Galactic star with 
a typical X-ray luminosity of $L_X$ $\sim$ 10$^{28}$$-$10$^{30}$ 
ergs s$^{-1}$, this object {\it must} be nearby with a distance of 
$\sim$30 $-$ 300 pc, and a V magnitude of $\sim$11 $-$ 14 is 
expected from the measured X-ray flux. We have, however, found no 
such bright stars in Galactic star catalogs. We find a faint star 
in the Guide Star Catalog with an $\sim$2$\arcsec$ offset from the 
source position, but the cataloged B magnitude is only 19.44. The 
best-fit absorbing column from the X-ray spectral modeling may also 
be higher ($N_H$ = 1.1 $\times$ 10$^{21}$ cm$^{-2}$) than the Galactic 
absorption toward the position of this hard X-ray knot ($N_H$ $\sim$ 
6 $\times$ 10$^{20}$ cm$^{-2}$; Dickey \& Lockman 1990). 

This hard X-ray knot is therefore most likely a feature spatially 
associated with SNR N49. This feature then appears to be fragments from 
the SN explosion ejected outside of the SNR shock boundary or 
``shrapnel'' as discovered in other Galactic SNRs such as Vela and Tycho 
\citep{aschenbach95,vancura95,warren02}. The true angular size and shape 
of this hard X-ray knot is however uncertain because of the poor off-axis 
PSF. We note that an on-axis {\it Chandra}/ACIS observation with a 
full angular coverage of N49, targeted on SGR 0526$-$66 has recently 
been performed (ObsID 2515). The detailed results from the diffuse
X-ray emission of the SNR have yet to be reported.  

\subsubsection{\label{subsubsec:sgr} Point Source }

The spectrum from the point-like source is extracted from region
7 (Figure \ref{fig:fig6}). Due to the angular offset from the
aim point, the PSF at the position of the point source is 
significantly degraded ($\ga$2$\arcsec$), and we select a relatively 
large area (a circular region with a radius of 6$\arcsec$), which 
contains $\sim$14000 photons. Although the total count rate 
($\sim$0.4 counts s$^{-1}$) is relatively high, pileup is 
insignificant due to the large off-axis angle. Since the source 
location is within fairly bright diffuse emission from the SNR,
the degraded PSF also makes the background estimation difficult.
We thus include a thermal component for the background spectrum 
in the spectral fitting instead of subtracting the background
from the source spectrum. 

\begin{figure}[]
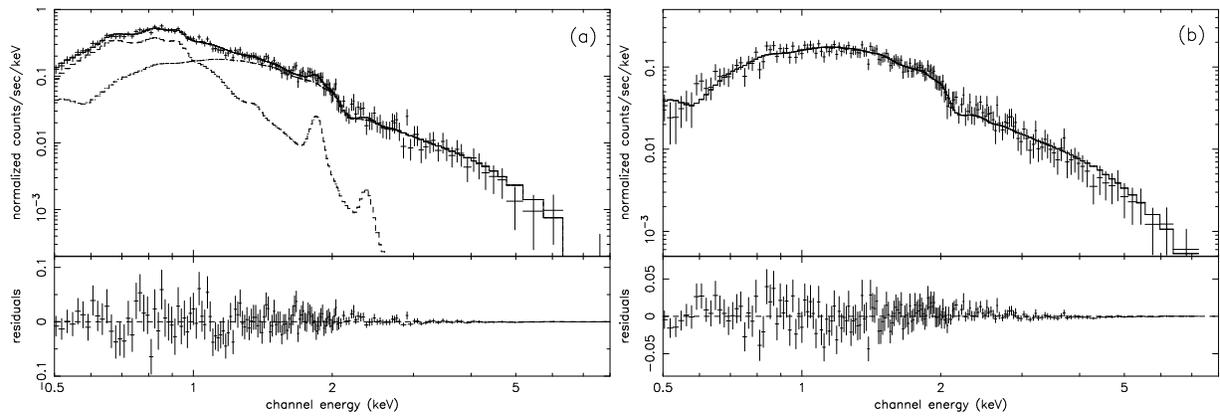

\figurenum{8}
\centerline{\includegraphics[angle=-90,width=8cm]{fig8a.ps}
\includegraphics[angle=-90,width=8cm]{fig8b.ps}}
\figcaption[]{ X-ray spectrum of SGR 0526$-$66 from {\it Chandra}/ACIS
observations. (a) is the spectrum obtained from the GTO data (region
7 in Figure \ref{fig:fig6}), and (b) is the spectrum extracted from 
the archival GO data. For the off-axis data, the background spectrum 
was included in the fitting as displayed with a soft thermal component
in (a).
\label{fig:fig8}}
\end{figure}

In a direct contrast to the diffuse emission spectra, the point 
source spectrum is dominated by continuum, as perhaps expected 
from a neutron star (Figure \ref{fig:fig8}a). The source spectrum 
can be described with a single power law model of photon index 
$\Gamma$ = 2.98$^{+0.17}_{-0.20}$. The embedded background thermal 
component is a $kT$ $\sim$ 0.33 keV plasma with LMC-like metal 
abundances, which is primarily responsible for the soft emission 
at $E$ $<$ 1 keV. 
A black body model may also fit the data ($kT$ $\sim$ 0.52 keV) 
with acceptable statistics, but appears, qualitatively, inappropriate 
to describe the hard tail of the specrum.  

Since the archival data are pointed in the direction of the position
of the point source or SGR 0526$-$66, the on-axis high angular 
resolution allows us to manipulate the background subtraction with 
less confusion. We extract the source spectrum from a circular 
region with a radius of 2$\arcsec$, which contains $\sim$10000 
photons. Because of the utilized short time-frame (0.4 s), pileup
is negligible with this on-axis observation. SGR 0526$-$66 is on 
top of complex filamentary structure of the diffuse emission from the 
SNR and the background emission for the SGR is non-uniform (Figure 
\ref{fig:fig1}b). We thus extract the background emission from the 
immediate annular region centered on the point source position (inner 
radius of 2$\farcs$5 and outer radius of 4$\arcsec$), in order to 
obtain an ``average'' spectrum over the variable emission features around 
the SGR. The SGR spectrum from the on-axis data can be fitted with 
a single power law model (Figure \ref{fig:fig8}b). The best-fit photon 
index is $\Gamma$ = 2.72$^{+0.11}_{-0.10}$ which is in good agreement 
with the result from the off-axis data. 
The results from the fit with a black body model ($kT$ $\sim$ 0.50 
keV) are also identical to the off-axis data and appear inadequate
by the reason of the same hard-tail discrepancy.
The implied X-ray luminosity is $L_X$ = 2.1 $\times$ 10$^{35}$ ergs 
s$^{-1}$ in the 2 $-$ 10 keV band with the off-axis data ($L_X$ = 
2.5 $\times$ 10$^{35}$ ergs s$^{-1}$ based on the on-axis data), 
which is consistent with the ASCA upper limit \citep{hughes98} and 
that of the Galactic SGR 1806$-$20 \citep{mereghetti00}. The 
results from the spectral model fits of the point source are
summarized in Table \ref{tbl:tab5}.

\section{\label{sec:disc} DISCUSSION}

\subsection{\label{subsec:shock} Multi-Phase Shock Structure}

The X-ray color images and the EW images indicate complex spectral 
structure of the diffuse X-ray emission across the SNR N49. The 
filamentary and clumpy emission features in the eastern half of 
the SNR show spatially distinctive bright structures in different 
colors. The spatial and spectral structure of the X-ray emission 
in the western half is relatively uniform, primarily represented
by intermediate-to-hard emission (green-blue) with low surface 
brightness. The soft (red-yellow) X-ray shell-like emission
features represent the blast wave shock front as observed in the
radio band. The overall features of the EW images indicate 
enhancements of soft line emission ($E <$ 1 keV) in the south-east 
and hard emission lines ($E >$ 1 keV) in the west of the SNR. 
These features are consistent with the overall density distributions 
of the ambient ISM; i.e., the blast wave is propagating into a low 
density ISM in the west and interacting with dense clouds in the east. 
The comparisons between the X-ray and optical images also suggest 
that the hard X-ray emission ($E >$ 1.6 keV) appears to be produced 
in less-dense intercloud regions while the soft X-rays ($E <$ 0.75 
keV) originate near the dense clouds. 

The results from the spatially-resolved spectral analysis confirm
the speculations inferred from the image analysis by revealing 
that shock temperatures and ionization states in the east of the 
SNR are significantly different from those in the western half: 
i.e., the western portion of the SNR is described by a simple 
one-temperature plane-parallel shock with relatively large 
ionization timescales while two-temperature model with multiple 
ionization timescales is required in the eastern half. The global 
spectral structure of the SNR may then be represented by a 
multi-phase shock model with three characteristic temperatures, 
i.e., the ``cool'' ($kT$ $\sim$ 0.44$^{+0.19}_{-0.10}$ keV) and 
``hot'' ($kT$ $\sim$ 1.25$^{+2.58}_{-0.22}$ keV) components in the 
east, and the ``intermediate'' ($kT$ $\sim$ 0.59$^{+0.09}_{-0.13}$ keV) 
component in the west. The presence of multiple phases of the shock 
due to the interactions of the blast wave with the dense molecular 
clouds has been put forth by Hester \& Cox (1986). Recently, based on 
the high resolution {\it Chandra}/ACIS observation of the Galactic SNR
Cygnus Loop, Levenson et al. (2002) reported evidence for the three 
distinct regions characterised as: (i) the forward shock decelerated 
by the clouds, (ii) the postshock SNR interior, and (iii) the reflected 
shock that re-heated the postshock regions. In this context, we propose 
that the cool component of X-ray emission from N49 corresponds to the 
forward shock entering the dense clouds, the intermediate component 
to emission from the postshock region, and the hot component to emission 
from the re-heated postshock region, respectively. Comparisons of the 
X-ray emission features with optical images indicate that the soft 
X-rays are spatially coincident with the bright optical knots while the 
hard X-ray emission is anti-correlated with the optical intensity 
distribution. Since the bright filamentary optical emission features 
are believed to originate from the slow, cool radiative shock entering 
dense molecular clouds, the observed correlation/anti-correlation 
between X-ray and optical images provide good support for the suggested 
multi-phase shock interpretation. 

Hester et al. (1994) have developed a simple 1-dimensional model 
in order to describe such a multi-phase shock structure of the nearby 
Galactic SNR Cygnus Loop, probably a ``magnified'' version of the
distant SNR N49. Our best-fit electron temperatures indicate that 
the ratios of temperatures between the re-heated postshock and the 
postshock without re-heating are $T_{rh}/T_{ps}$ = $\sim$1.8 $-$ 2.3, 
which implies the preshock cloud density $n_c$ $\sim$ 30 $-$ 2500 $n_0$ 
based on the model by Hester et al. (1994), where $n_0$ is the preshock 
intercloud density. Assuming the preshock intercloud density $n_0$ $\sim$ 
0.9 cm$^{-3}$ as estimated from the optical spectroscopy \cite{vancura92}, 
we derive $n_c$ $\sim$ 27 $-$ 2300 cm$^{-3}$ for the interacting 
dense molecular clouds. This is in plausible agreements with the 
previous estimations based on optical/UV spectroscopy ($n_c$ = 20 $-$ 
940 cm$^{-3}$; Vancura et al. 1992). 
The best-fit $EM$s in the eastern regions imply 
($n_c$/$n_0$)$^2$($V_c$/$V_{rh}$) = 2 $-$ 13, where $V_c$ is the X-ray 
emission volume of the cool forward shock component and $V_{rh}$ 
is the X-ray emission volume of the hot re-heated component (we assume
an X-ray emitting volume filling factor of $\sim$1, hereafter). Provided 
that $n_c$/$n_0$ is in the order of $\sim$10$^{1}$ $-$ 10$^{3}$, this 
implies $V_c$/$V_{rh}$ $\la$ 0.01. The soft X-ray emission from the 
cool forward shock is then from the blast wave entering small-volume, 
clumpy clouds while the hot component originates 
from the large-volume, re-heated intercloud postshock regions. 
For an adiabatic shock, the relations between the shock temperature 
and the shock velocity is $T_s$ = $\frac{3\bar{m}v_s^2}{16k}$ and 
can be written as $$kT_s = 0.013(v_s/100~km~s^{-1})^{2}~keV,$$
where $T_s$ is the shock temperature, $v_s$ is the shock velocity,
and mean atomic mass $\bar{m}$ $\approx$ 0.7$m_p$ ($m_p$ = proton 
mass). The cool component shock velocity is then $v_c$ = 582$^{+114}_{-71}$ 
km s$^{-1}$, the hot component velocity is $v_h$ = 981$^{+735}_{-91}$ 
km s$^{-1}$, and the intermediate velocity is $v_i$ = 674$^{+49}_{-79}$ 
km s$^{-1}$. For comparisons, Vancura et al. (1992) have derived a
shock velocity of $v$ = 730 km s$^{-1}$ for the X-ray emission of N49
(Note: For simplicity, we assumed the electron and ion temperatures
in equalibration in these calculations).

The ionization timescale ($n_et$) ratios between the cool and the
hot components are ($n_et$)$_c$/($n_et$)$_{rh}$ $\la$ 0.1, 
which implies $t_c$ $\la$ 0.01$t_{rh}$, where $t_c$ is the
ionization age of the shock entering the clouds and $t_{rh}$ is the
ionization age of the re-heated regions. Assuming an SN explosion 
energy of $E_0$ = 10$^{51}$ ergs, the SNR radius of $r$ = 10 pc 
based on the X-ray image of N49, and the preshock density $n_0$ = 
0.9 cm$^{-3}$, the implied Sedov dynamical age \citep{sedov59}, 
$t_d$ is $\sim$ 6600 yr. Assuming $t_{rh}$ is on the order of 
$t_d$ or $\sim$10$^{4}$ yr, the measured $n_et$ ratios imply 
$t_c$ $\la$ 100 yr. This short ionization age suggests that the 
soft X-ray emission ($kT$ $\sim$ 0.44 keV) has been produced by 
a rapidly decelerated shock entering dense molecular 
clouds as was observed from the Cygnus Loop SNR \citep{hester94}.

\subsection{\label{subsec:bullet} Supernova Ejecta Fragments}

The hard X-ray knot in the southwest beyond the SNR blast wave 
shock front (region 6; Figure \ref{fig:fig6}) is most likely 
associated with the SNR rather than a background/foreground object. 
The X-ray spectrum is dominated by emission lines and can be 
described with a $kT$ $\sim$ 1 keV plasma with metal-rich 
abundances, which supports the shocked ejecta origin. The fitted 
plane-parallel shock model indicates a highly advanced ionization 
state ($n_et$ $\sim$5 $\times$ 10$^{13}$ cm$^{-3}$ s$^{-1}$) and 
the spectrum may also be equally described with a thermal plasma 
in the CIE. The high temperature and the large ionization timescale 
of this feature are also supportive of its origin as the high-speed 
ejecta fragments created at the time of the SN explosion. 

Because of the PSF degradation, the apparent angular extension of 
this feature ($\sim$5$\arcsec$ in radius) is likely an artifact and 
the true morphology is uncertain with the current data. In order to
estimate the electron density $n_e$, we thus assume two extreme 
cases of a circular shape with a radius of 5$\arcsec$ (for the 
apparent angular size and so 1.2 pc radius in physical size
at the distance of $d$ = 50 kpc) and 0$\farcs$5 (for the case 
of an unresolved point-like source, and so the physical size of 
$\la$ 0.125 pc radius at $d$ = 50 kpc). The assumed range of the 
physical size of the knot ($\sim$0.1 pc $-$ 1.2 pc in radius) is 
reasonably consistent with the SN ejecta fragments detected in 
other Galactc SNRs of Vela and Tycho \citep{vancura95,miyata01}. 
Assuming the distance of 50 kpc to the SNR and a simple spherical 
geometry for the X-ray emitting volume, the best-fit $EM$ implies 
$n_e$ of 2.3 cm$^{-3}$ to $\ga$ 68 cm$^{-3}$ depending on the 
assumed emission volume. Based on the best-fit abundances, we 
may derive the ejecta mass of $\sim$0.001 $M_{\odot}$ $-$ 
0.03 $M_{\odot}$. Although this range of the ejecta mass is in 
reasonable agreements with other Galactic SN ejecta fragments
\citep{tsunemi99,aschenbach95,strom95}, we caution that these 
estimations are crude due to the large uncertainties in the 
fitted abundances and assumed emission volumes.  

The angular offset of the hard X-ray knot from the geometrical
center of N49 is $\sim$48$\arcsec$. The spatial separation of
this knot from the center of the SNR is then $D$ = 11.6 $d_{50}$ 
($\sin\theta$)$^{-1}$ pc, where $d_{50}$ is the distance to N49 in 
the units of 50 kpc, and $\theta$ is the angle between the line of 
sight and the moving direction of the knot. Assuming that the SN
explosion site, so the birth place of the observed ejecta fragments,
is close to the geometrical center of the current X-ray remnant
of N49, the mean transverse velocity is $v_{mean}$ = 1717 $D_{11.6}$ 
$t^{-1}_{6600}$ km s$^{-1}$, where $D_{11.6}$ is the spatial 
separation of the knot from the geometrical center of the SNR in 
the units of 11.6 pc, and $t_{6600}$ is the age of the SNR in the 
units of 6600 yr. 
The analysis on this hard X-ray knot is however substantially limited 
by the poor PSF. Follow-up on-axis {\it Chandra} observations would be 
essential to unveil the detailed nature of this feature. 

\subsection{\label{subsec:point} SGR 0526$-$66}

The {\it Chandra}/ACIS observations conclusively detect an X-ray
point source at the position of SGR 0526$-$66 within the boundary
of SNR N49. The X-ray spectrum of this point source is dominated 
by continuum, which can be described with a power law model (see
\S \ref{subsubsec:sgr}). This point source is thus confidently 
identified with the X-ray counterpart of quiescent phase SGR 0526$-$66. 

The single power law model indicates a soft spectrum with $\Gamma$ 
$\sim$ 3 for SGR 0526$-$66, which resembles the typical spectrum of
the AXPs (Mereghetti et al. 2002 and references therein), while the 
X-ray spectra from Galactic SGRs are described with harder single 
power laws of $\Gamma$ $\sim$ 2 
\citep{sonobe94,kouveliotou98,woods99,mereghetti00,kouveliotou01}.
As X-ray spectra from the AXPs and SGRs can be described with the
power law + black body models, we have made an attempt to fit the 
SGR 0526$-$66 spectrum with such a multi-component model. With
the off-axis GTO data, the fitted black body temperature ($kT$ 
$\sim$ 0.55 keV) is consistent with typical AXP/SGR spectra, and 
the improvement in the overall fit is also statistically significant 
based on the F-test. The power law component, with the addition
of the black body component, however implies substantially harder 
spectrum ($\Gamma$ $\sim$ 1.7) than those of the AXPs ($\Gamma$ 
$\sim$ 3$-$4; Mereghetti et al. 2002 and references therein). 
Nonetheless, the two component model is still in good agreements 
with Galactic SGRs as they are fitted with $\Gamma$ $\sim$ 1$-$2
and black body temperature $kT$ $\sim$ 0.5 keV 
\citep{woods99,kouveliotou01}. We however note that this two 
component model is not required to describe the on-axis data.
Since the embedded confusion level caused by the background in 
the spectral model fitting is more severe for the off-axis data 
than the on-axis data, we consider that the presence of the black 
body component is inconclusive, although the improvement in the 
fit for the off-axis data is statistically significant with the 
additional black body component. Recently, Kulkarni et al. (2002)
have reported the results from the spectral analysis of SGR 0526$-$66
utilizing two {\it Chandra}/ACIS GO observations: one (ObsID 747) 
of those two data sets was identical to that we uitilze in the current 
work. They found that the observed spectrum of the SGR was best described 
by a black body + power law model, which appears to be inconsistent 
with our results. Although the origin of the inconsistency is unclear, 
we may attribute the descrepancy to the background estimations utilized
in the spectral analysis. As Kulkarni et al. (2002) pointed out, their 
SGR source spectrum apparently include residual emission from the 
diffuse SNR, whereas such a contamination in our on-axis source spectrum
appears to be negligible. The soft X-rays from the diffuse emission 
of the SNR might have caused the slight differences in the results
of the spectral analysis as presented by Kulkarni et al. (2002). We have 
also noticed that Kulkarni et al. (2002) have not incorperated the CTI 
and QE corrections in their data reduction. Although the substantial 
contributions by the CTI and the QE degradation effects on a point-like 
source in the observation taken in a relatively early phase of the
{\it Chandra} mission are unlikely, we speculate that potential
contaminations from these artifacts in the spectral analysis of any
ACIS data cannot be completely ruled out. In addition, we would like
to comment on the spectral analysis of some parts of the diffuse emission
from N49 by Kulkarni et al. (2002). They have used a MEKAL model for
such a spectral analysis, which should be inappropriate for the 
complicated diffuse SNR emission as we presented in the current work. 
More complex, physically realistic models should be utilized in order 
for an appropriate spectral analysis of such data.

The results from our simple spectral analysis may support the 
speculation that SGR 0526$-$66 is a transition object bridging 
the AXP and the SGR phases of neutron stars \citep{kaplan01}.
Extensive analyses with the on-axis observations, which are beyond 
the scope of the current work, would however be necessary to 
understand the detailed astrophysical characteristics of this 
object. For example, the measurements of the period ($P$) and the 
period derivative ({\it \.{P}}) would be critical to determine the 
magnetic field strengh $B$ and the spin-down age $\tau$ of SGR 
0526$-$66. This information can provide important clues on the 
nature of SGR 0526$-$66 such as the identification with a high-field 
($B$ $\ga$ 10$^{15}$ G) neutron star or the ``magnetar'' 
\citep{paczynski92,duncan92}. The measurement of the spin-down age 
would also be useful to test the physical association of SGR 0526$-$66 
with SNR N49, which should provide important implications on the 
possible astrophysical connections between the SGRs and the AXPs 
(e.g., Gaensler et al. 2001). Kulkarni et al. (2002) have recently
reported a detection of the 8.04 s pulsation from SGR 0526$-$66 by 
utilizing $\sim$85 ks {\it Chandra}/ACIS observations. The derived 
magnetic dipole field strength is high ($B$ $\sim$ 7 $\times$ 10$^{14}$ G) 
and supports the magnetar interpretation. The estimated spin-down 
age ($P$/2{\it \.{P}} $\sim$ 2000 yr) is consistent with the dynamical 
age of SNR N49, which support the physical association between N49
and SGR 0526$-$66. The location of SGR 0526$-$66 in the LMC and
the likely association with N49 is also supported by the fitted columns:
i.e., the absorbing columns for N49 and SGR 0526$-$66 are consistent.
This SGR location may confirm the enormous energy release in the
1979 March $\gamma$-ray burst.

If SGR 0526$-$66 is indeed the compact remnant of the SN explosion
which has also created SNR N49 and it was born near the center of N49,
we may estimate its transverse velocity. The angular offset of SGR 
0526$-$66 from the geometrical center of N49 is $\sim$22$\arcsec$, 
which implies that the spatial separation of the SGR from the center 
of the SNR is $D$ = 5.3 ($\sin\theta$)$^{-1}$ pc at the distance of 
50 kpc. Assuming that SGR 0526$-$66 is the neutron star born at
the time of the SN explosion that produced N49, and that the age 
of the SGR is $\tau$ $\sim$ 6600 yr, the mean transverse velocity is 
$v_{mean}$ $\ga$ 787 km s$^{-1}$. This would suggest that SGR 0526$-$66 
is a young, high velocity neutron star that is in a different class 
than typical AXPs instead of being the same class object as other AXPs, 
simply in a more evolved phase. 

\section{\label{sec:summary} SUMMARY}

SNR N49 in the LMC has been observed with {\it Chandra}/ACIS.
The high resolution ``true-color'' ACIS images reveal complex
X-ray filamentary structures which represent the blast wave shock
sweeping through the ambient ISM and the shock interacting with the
nearby dense molecular clouds. The overall features from the EW
images of the detected elemental species indicate that soft line 
emission is enhanced in the south-east while hard line emission
is enhanced in the west of the SNR. These overall EW features are 
most likely dominated by the variations in the shock temperature 
and the ionization states of the plasma across the SNR. The 
spatially-resolved spectral analysis of N49 suggests that the 
X-ray emission from the SNR can be described with multiple-phase 
shocks representing the singly-postshock region, the re-heated 
postshock region by the reflected shocks, and the cool forward 
shock entering the dense molecular clouds.

We detect a hard X-ray knot in the southwest of N49 beyond
the SNR shock boundary. The X-ray spectrum of this knot is emission
line dominated and can be described with a metal-rich hot thermal
plasma. This result implies that this hard knot appears to be
ejecta fragments created at the time of the SN explosion that
produced SNR N49. The PSF degradation of the off-axis observation 
however restricts the data analysis of this feature and preclude
resolving the angular structure of this knot. Follow-up on-axis 
observations with a decent exposure might be necessary to address 
the detailed nature of this feature.

The ACIS observations of N49 also firmly detect the X-ray counterpart
of the quiescent phase of SGR 0526$-$66. The X-ray spectrum of SGR 
0526$-$66 is continuum dominated and can be fitted with a power law 
model. The implied photon index is $\Gamma$ $\sim$ 3. This power law 
spectrum is generally consistent with those from the AXPs although the 
observed $\gamma$-ray burst of SGR 0526$-$66 is a rare characteristic 
of typical AXPs. 

\acknowledgments

The authors thank J. Dickel for providing his ATCA data. SP also thank 
A. Bykov for a valuable discussion on the SN ejecta fragments.
This work has been in parts supported by NASA under contract
NAS8-01128 for the {\it Chandra X-Ray Observatory}. JPH acknowledges 
support from {\it Chandra} grants GO1-2052X, GO2-3068X, and GO2-3080B 
to Rutgers.

\clearpage

\begin{deluxetable}{cccc}
\footnotesize
\tablecaption{Energy Bands used for Generating the Equivalent Width
Images.
\label{tbl:tab1}}
\tablewidth{0pt}
\tablehead{ \colhead{Elements} & \colhead{Line} &
\colhead{Low\tablenotemark{a}} & \colhead{High\tablenotemark{a}} \\
 & \colhead{(eV)} & \colhead{(eV)} & \colhead{(eV)} }
\startdata
O & 590 $-$ 740 & 300 $-$ 550 & 1200 $-$ 1280 \\
Fe L & 750 $-$ 870 & 300 $-$ 550 & 1200 $-$ 1280 \\
Ne (He$\alpha$) & 880 $-$ 950 & 300 $-$ 550 & 1200 $-$ 1280 \\
Ne (Ly$\alpha$)/Fe L & 980 $-$ 1100 & 300 $-$ 550 & 1200 $-$ 1280 \\
Mg & 1290 $-$ 1550 & 1200 $-$ 1280 & 1630 $-$ 1710 \\
Si & 1750 $-$ 2100 & 1630 $-$ 1710 & 2140 $-$ 2280 \\
\enddata
\tablenotetext{a}{The high and low energy bands around the
selected line energies used to estimate the underlying continuum.}

\end{deluxetable}

\clearpage

\begin{deluxetable}{ccccccc}
\footnotesize
\tablecaption{Results from the Spectral Model Fittings: Western Regions
of N49\tablenotemark{a}.
\label{tbl:tab2}}
\tablewidth{0pt}
\tablehead{ \colhead{Region\tablenotemark{b}} & \colhead{$N_H$} &
 \colhead{$kT$} & \colhead{$n_et$\tablenotemark{c}} & 
\colhead{$EM$\tablenotemark{d}} & \colhead{$\chi^{2}$} & \colhead{$\nu$} \\
 & \colhead{(10$^{21}$ cm$^{-2}$)} & \colhead{(keV)} & 
\colhead{(10$^{11}$ cm$^{-3}$ s)} & \colhead{(10$^{58}$ cm$^{-3}$)} & &}
\startdata
1& 2.7$^{+1.0}_{-1.1}$ & 0.52$^{+0.13}_{-0.11}$ & 
8.6$^{+5.3}_{-3.2}$ & 2.01$^{+0.84}_{-0.90}$ & 43.3 & 60 \\
2 & 1.2$^{+0.6}_{-0.5}$ & 0.65$^{+0.04}_{-0.05}$ &
70.0$^{+104.0}_{-30.0}$ & 2.79$^{+0.72}_{-0.75}$ & 77.0 & 93 \\
3 & $<$ 2.5 (0.6) & 0.59$^{+0.05}_{-0.18}$ & 
31.5$^{+108.5}_{-19.0}$ & 1.71$^{+0.51}_{-0.51}$ & 59.9 & 56 \\ 
6 & $<$ 3.6 (1.1) & 1.04$^{+0.11}_{-0.09}$ &
$>$ 41.0 (500) & $<$ 0.27 (0.12) & 19.5 & 36 \\
\enddata
\tablenotetext{a}{Errors are with a 90\% confidence. The 90\%
limits (with the best-fit values in the parentheses) are presented
where the best-fit parameters are unconstrained.}
\tablenotetext{b}{Regions as marked in Figure \ref{fig:fig6}.}
\tablenotetext{c}{Errors are obtained after fixing $N_H$ and $kT$
at the best-fit values.}
\tablenotetext{d}{Volume emission measure, $EM$ = ${\int}n_en_HdV$.}
\end{deluxetable}

\clearpage

\begin{deluxetable}{ccccccc}
\footnotesize
\tablecaption{Fitted Metal Abundances: Western Regions of N49\tablenotemark{a}.
\label{tbl:tab3}}
\tablewidth{0pt}
\tablehead{ \colhead{Region\tablenotemark{b}} & \colhead{O} &
 \colhead{Ne} & \colhead{Mg} & \colhead{Si} & \colhead{S} & \colhead{Fe} }
\startdata
1 & 0.90$^{+0.83}_{-0.30}$ & 1.01$^{+0.98}_{-0.36}$ & 
0.53$^{+0.62}_{-0.25}$ & 0.39$^{+0.54}_{-0.27}$ & $<$ 5.77 (1.91) & 
0.17$^{+0.16}_{-0.06}$ \\
2 & 1.23$^{+0.61}_{-0.34}$ & 1.80$^{+0.84}_{-0.47}$ &
0.78$^{+0.42}_{-0.24}$ & 0.86$^{+0.41}_{-0.25}$ & 1.75$^{+1.36}_{-0.99}$ &
0.25$^{+0.10}_{-0.06}$ \\
3 & 0.46$^{+0.29}_{-0.17}$ & 0.30$^{+0.26}_{-0.17}$ &
0.16$^{+0.19}_{-0.12}$ & 0.19$^{+0.24}_{-0.19}$ & $<$ 4.02 (1.50) &
0.18$^{+0.09}_{-0.05}$ \\
6 & 5.71$^{+40.2}_{-4.16}$ & 0.33 (fixed) &
$<$ 502.7 (1.11) & 5.92$^{+34.0}_{-3.72}$ & $<$ 1334.2 (3.98) &
0.94$^{+1.46}_{-0.61}$ \\
\enddata

\tablenotetext{a}{The 90\% errors have been estimated with 
$N_H$, $kT$, and $n_et$ fixed at the best-fit values. The 90\%
limits (with the best-fit values in the parentheses) are presented
where the best-fit parameters are unconstrained.}

\tablenotetext{b}{Regions as marked in Figure \ref{fig:fig6}.}

\end{deluxetable}

\clearpage

\begin{deluxetable}{cccccccc}
\footnotesize
\tablecaption{Results from the Spectral Model Fittings: Eastern Regions
of N49\tablenotemark{a}.
\label{tbl:tab4}}
\tablewidth{0pt}
\tablehead{\colhead{Region\tablenotemark{b}} & \colhead{$N_H$} &
\colhead{$\chi^{2}$} & \colhead{$\nu$} & Shock &\colhead{$kT$} & 
\colhead{$n_et$\tablenotemark{c}} & \colhead{$EM$\tablenotemark{d}} \\
 & \colhead{(10$^{21}$ cm$^{-2}$)} & & & Components &\colhead{(keV)} &
\colhead{(10$^{11}$ cm$^{-3}$ s)} & \colhead{(10$^{58}$ cm$^{-3}$)} }
\startdata
4& 3.9$^{+1.2}_{-2.6}$ & 67.7 & 94 & soft & 0.47$^{+0.24}_{-0.12}$ & 
0.8$^{+0.7}_{-0.3}$ & 3.81$^{+4.29}_{-3.78}$ \\
 & & & & hard & 1.08$^{+2.32}_{-0.12}$ & $>$ 14.6 (40.9) &
2.19$^{+0.84}_{-1.59}$ \\
5 & 3.6$^{+0.8}_{-1.1}$ & 86.7 & 96 & soft & 0.41$^{+0.13}_{-0.08}$ & 
1.1$^{+0.5}_{-0.4}$ & 19.83$^{+5.58}_{-3.75}$  \\
 & & & & hard & 1.41$^{+2.79}_{-0.31}$ & $>$ 9.3 (22.1) &
1.53$^{+0.57}_{-1.50}$ \\
\enddata
\tablenotetext{a}{Errors are with a 90\% confidence. }
\tablenotetext{b}{Regions as marked in Figure \ref{fig:fig6}.}
\tablenotetext{c}{Errors are obtained after fixing $N_H$ and $kT$
at the best-fit values. The 90\% limits (with the best-fit values
in the parentheses) are presented where the best-fit parameters 
are unconstrained.}
\tablenotetext{d}{Volume emission measure, $EM$ = ${\int}n_en_HdV$.}
\end{deluxetable}

\clearpage

\begin{deluxetable}{cccccc}
\footnotesize
\tablecaption{Results from the Spectral Model Fittings: SGR
0526$-$66\tablenotemark{a}.
\label{tbl:tab5}}
\tablewidth{0pt}
\tablehead{\colhead{Region\tablenotemark{b}} & 
\colhead{$N_H$\tablenotemark{c}} & \colhead{$\Gamma$\tablenotemark{c}} &
\colhead{$L_X$\tablenotemark{d}} & \colhead{$\chi^{2}$} & \colhead{$\nu$} \\
 & \colhead{(10$^{21}$ cm$^{-2}$)} & & \colhead{(10$^{35}$ ergs s$^{-1}$)} & &}
\startdata
 7 & 4.4$^{+0.8}_{-1.0}$ & 2.98$^{+0.17}_{-0.20}$ & 2.1 & 113.7 & 143\\
on-axis & 4.7$^{+0.4}_{-0.3}$ & 2.72$^{+0.11}_{-0.10}$ & 2.5 & 131.8 & 170 \\
\enddata
\tablenotetext{a}{The results of the spectral fits with a power law model
from both of our GTO data and the archival GO data are presented.}
\tablenotetext{b}{Region \#, as marked in Figure \ref{fig:fig6}, is
presented for the GTO data. ``On-axis'' indicates the source region
from the archival data.}
\tablenotetext{c}{Errors are with a 90\% confidence. }
\tablenotetext{d}{The 2 $-$ 10 keV band X-ray luminosity assuming
a distance of 50 kpc to the source.}
\end{deluxetable}

\clearpage

\clearpage

\clearpage

\clearpage

\clearpage

\clearpage

\clearpage

\clearpage

\end{document}